\documentclass[conference]{IEEEtran}
\IEEEoverridecommandlockouts
\usepackage{cite}
\usepackage{amsmath,amssymb,amsfonts}
\usepackage{graphicx}
\usepackage{textcomp}
\usepackage{xcolor}
\usepackage{multirow}
\usepackage{algorithm}
\usepackage{algpseudocode}
\usepackage{tikz}
\usetikzlibrary{fit,positioning}
\usepackage{enumitem}
\usepackage{hyperref}
\usepackage{booktabs}       
\usepackage{multirow}       
\usepackage{array}          
\usepackage{colortbl}       

\usepackage{xcolor}         
\usepackage{pifont}         

\newcommand{\cmark}{\textcolor{green!60!black}{\ding{51}}} 
\newcommand{\xmark}{\textcolor{red!70!black}{\ding{55}}}   

\def\BibTeX{{\rm B\kern-.05em{\sc i\kern-.025em b}\kern-.08em
    T\kern-.1667em\lower.7ex\hbox{E}\kern-.125emX}}

\renewcommand{\baselinestretch}{0.9}

\begin{document}

\title{DORA: Dynamic O-RAN Resource Allocation for Multi-Slice 5G Networks\\

\thanks{This material is based upon work supported by the National Science Foundation under Grant Numbers  CNS-2202972, CNS-2232048, and CNS-2318726.}
}

\author{
	\IEEEauthorblockN{
        Alireza Ebrahimi Dorcheh\IEEEauthorrefmark{1}, 
	Tolunay Seyfi\IEEEauthorrefmark{1}, 
        Fatemeh Afghah\IEEEauthorrefmark{1}}

    \IEEEauthorblockA{\IEEEauthorrefmark{1}Holcombe Department of Electrical and Computer Engineering, Clemson University, Clemson, SC, USA \\
        Emails: \{alireze, tseyfi,  
        fafghah\}@clemson.edu}

}



\maketitle

\begin{abstract}
The fifth generation (5G) of wireless networks must simultaneously support heterogeneous service 
categories, including Ultra-Reliable Low-Latency Communications (URLLC), enhanced Mobile Broadband 
(eMBB), and massive Machine-Type Communications (mMTC), each with distinct Quality of Service (QoS) 
requirements. Meeting these demands under limited spectrum resources requires adaptive and 
standards-compliant radio resource management. We present DORA (Dynamic O-RAN Resource Allocation), 
a deep reinforcement learning (DRL) framework for dynamic slice-level Physical Resource Block (PRB) 
allocation in Open RAN. DORA employs a PPO-based RL agent to allocate PRBs across URLLC, eMBB, 
and mMTC slices based on observed traffic demands and channel conditions. Intra-slice PRB scheduling 
is handled deterministically via round-robin among active UEs, simplifying control complexity and 
improving training stability. Unlike prior work, DORA supports online training and adapts 
continuously to evolving traffic patterns and cross-slice contention.
Implemented in the standards-compliant OpenAirInterface (OAI) RAN stack and designed for deployment 
as an O-RAN xApp, DORA integrates seamlessly with RAN Intelligent Controllers (RICs). Extensive 
evaluation under congested regimes shows that DORA outperforms three non-learning baselines and a 
\texttt{DQN} agent, achieving lower URLLC latency, higher eMBB throughput with fewer SLA violations, 
and broader mMTC coverage without starving high-priority slices. To our knowledge, this is the first 
fully online DRL framework for adaptive, slice-aware PRB allocation in O-RAN.
\end{abstract}


\section{Introduction}
\label{sec:introduction}

The fifth generation (5G) of wireless networks introduces heterogeneous service categories, 
including enhanced Mobile Broadband (eMBB), Ultra-Reliable Low-Latency Communication (URLLC), 
and massive Machine-Type Communication (mMTC), each with distinct Quality of Service (QoS) 
requirements. Meeting these diverse demands under limited spectrum resources necessitates adaptive and intelligent radio resource management strategies.

The Open Radio Access Network (O-RAN) paradigm addresses this challenge by disaggregating the 
RAN into modular components connected via open interfaces, enabling vendor-neutral innovation 
and real-time control. Central to O-RAN are RAN Intelligent Controllers (RICs), which host 
xApps and rApps capable of leveraging AI and machine learning for dynamic spectrum sharing, 
mobility optimization, and PRB allocation. Experimental testbeds such as 
POWDER~\cite{JohnsonPOWDER2022}, COSMOS~\cite{raychaudhuri2020cosmos}, 
AERPAW~\cite{MooreAerpaw}, and Colosseum~\cite{polese2024colosseum} have accelerated 
O-RAN research by enabling realistic end-to-end experimentation combining terrestrial and aerial 
infrastructures.

Despite this progress, existing PRB allocation approaches face significant limitations.  
Static slicing strategies cannot adapt to dynamic traffic loads and channel variations, resulting 
in suboptimal QoS compliance. Many reinforcement learning (RL)-based solutions, while adaptive, 
are trained offline and fail to generalize to real-time environments with evolving cross-slice 
contention. Moreover, several frameworks jointly handle slice-level and per-UE scheduling, 
increasing control complexity and slowing convergence, while others assume simultaneous evaluation 
of all UEs, limiting scalability.

To address these challenges, we propose \textbf{DORA} (\textbf{D}ynamic \textbf{O}-RAN 
\textbf{R}esource \textbf{A}llocation), a deep reinforcement learning (DRL) framework for 
adaptive slice-level PRB allocation in O-RAN. DORA uses a PPO-based RL agent to 
dynamically allocate PRBs across URLLC, eMBB, and mMTC slices, optimizing service-level 
agreements (SLAs) under fluctuating traffic and channel conditions. Intra-slice PRB scheduling 
is handled deterministically via a round-robin policy among active UEs, reducing control 
complexity and enabling fast convergence.

Our framework is implemented within the standards-compliant OpenAirInterface (OAI) RAN stack and 
integrates seamlessly with FlexRIC, OAI's O-RAN-compliant RIC environment, enabling deployment as an xApp. By leveraging 
software based UEs, DORA supports scalable experimentation without physical hardware dependencies. 

This work makes the following key contributions:
\begin{itemize}[left=0pt]
    \item \textbf{Online DRL-based PRB allocation:} We introduce DORA, a single-agent PPO-based framework for adaptive slice-level PRB allocation under dynamic traffic and channel conditions.
    \item \textbf{O-RAN compliance and seamless integration:} DORA is implemented in the OpenAirInterface RAN stack and designed for deployment as an xApp via standard E2 interfaces.
    \item \textbf{Reduced control complexity:} Intra-slice PRB scheduling uses deterministic round-robin allocation, decoupling per-UE scheduling from slice-level optimization to improve scalability.
    \item \textbf{Extensive evaluation:} Using an OAI-based testbed, we show that DORA outperforms three non-learning baselines and a \texttt{DQN} agent, achieving lower URLLC latency, higher eMBB throughput, and broader mMTC coverage under congestion.
    \item \textbf{Scalable multi-UE evaluation:} DORA considers a total of 14 UEs across three standardized slices, that is 2 URLLC, 2 eMBB, and 10 mMTC, and supports fully online evaluation with up to four software-based UEs running in parallel within the OAI testbed, enabling realistic, high-fidelity performance assessment without requiring physical hardware.

\end{itemize}

To the best of our knowledge, DORA is the first online, standards-compliant DRL framework that enables adaptive, slice-aware PRB allocation in O-RAN, supporting realistic multi-UE evaluation across diverse traffic profiles.

\section{Related Work}
\label{sec:related_work}

Deep reinforcement learning (DRL) has emerged as a powerful approach for adaptive physical resource block (PRB) allocation and slice-aware resource management in O-RAN. 
PandORA~\cite{tsampazi2025pandora} introduces a framework for automating the design and benchmarking of DRL-based xApps, comparing 23 agents with varying action spaces, reward functions, and granularity levels. Similarly, Tsampazi \emph{et al.}~\cite{tsampazi2023comparative} evaluate 12 DRL-based xApps on Colosseum, highlighting how design choices affect performance under diverse network conditions. 
ColO-RAN~\cite{polese2022colo} addresses practical deployment challenges by developing three xApps for slicing, scheduling, and online training, illustrating the feasibility of adaptive allocation in realistic environments. OrchestRAN~\cite{doro2024orchestrran} complements these efforts by orchestrating where and how AI models are deployed across O-RAN components to satisfy latency and resource constraints.

The REAL framework~\cite{barker2025real} demonstrates an online PPO-based xApp integrated into an O-RAN-compliant stack using srsRAN and OSC RIC, achieving adaptive PRB allocation for up to 12 dynamic UEs. However, REAL simulates physical-layer processing using a GNU Radio-based PHY emulator, bypassing the detailed NR-compliant PHY/MAC stack. In contrast, DORA leverages the RFsim environment of the OAI stack, enabling high-fidelity simulation of 3GPP-standard PHY and MAC procedures. Furthermore, unlike REAL, which jointly handles slice- and UE-level scheduling within a single agent—leading to higher control complexity and reduced scalability—DORA decouples slice-level PRB allocation from per-UE scheduling, employing deterministic round-robin assignment within slices. This simplification improves convergence, reduces training complexity, and enhances adaptability under congestion, while supporting realistic evaluation with up to 14 UEs.

In parallel, several works focus on adaptive slice reconfiguration rather than per-TTI PRB allocation. 
AdaSlicing~\cite{zhao2025adaslicing} and InSlicing~\cite{zhao2025inslicing} employ optimization techniques and surrogate models to enhance control-plane flexibility and interpretability but do not handle intra-slice scheduling. 
ORANSlice~\cite{cheng2024oranslice} provides an open-source OAI-based implementation of 3GPP-compliant RAN slicing via the E2SM-CCC service model, serving as a programmable platform upon which intelligent PRB allocation xApps like DORA can operate.

Despite these advances, most existing frameworks either rely on offline data, assume simultaneous multi-UE evaluation, or tightly couple slice- and UE-level scheduling, limiting scalability under real-time constraints. 
DORA fills this gap by introducing the first standards-compliant DRL framework implemented in the OAI RAN stack for adaptive slice-level PRB allocation. Its key innovation lies in its fully online evaluation capability, supporting up to four UEs running simultaneously while training is performed offline. Extensive evaluation demonstrates that DORA outperforms three non-learning baselines and a \texttt{DQN} agent, achieving lower URLLC latency, higher eMBB throughput, and improved mMTC coverage under congestion.

\begin{table}[H]
    \centering
    \footnotesize
    \renewcommand{\arraystretch}{1.15}
    \setlength{\tabcolsep}{2.5pt} 
    \begin{tabular}{p{2.7cm}ccccc}
        \toprule
        \textbf{Framework} & \textbf{Onl.} & \textbf{Slice} &
        \textbf{UE Sch.} & \textbf{OAI/O-RAN} & \textbf{3GPP PHY} \\ \midrule
        \textbf{PandORA}~\cite{tsampazi2025pandora}   & \xmark  & \cmark & \cmark & \cmark & HW \\
        \textbf{ColO-RAN}~\cite{polese2022colo}       & Partial & \cmark & \cmark & \cmark & HW \\
        \textbf{REAL}~\cite{barker2025real}           & \cmark  & \cmark & \cmark & \cmark & \xmark \\
        \textbf{AdaSlicing}~\cite{zhao2025adaslicing} & \xmark  & \cmark & \xmark & \xmark & \xmark \\
        \textbf{ORANSlice}~\cite{cheng2024oranslice}  & \xmark  & \cmark & \xmark & \cmark & SW \\ \midrule
        \rowcolor{gray!10}
        \textbf{DORA} & \textbf{\cmark} & \textbf{\cmark} & \textbf{Det.} & \textbf{\cmark} & \textbf{SW} \\ 
        \bottomrule
    \end{tabular}
    \vspace{2pt}
    \caption{Comparison of DORA with representative O-RAN resource allocation frameworks. 
    DORA uses a software-based 3GPP PHY via OAI RFsim, while PandORA and ColO-RAN rely on 
    hardware-based PHY. REAL uses a GNU Radio PHY emulator.}
    \label{tab:framework_comparison}
\end{table}

\section{System Model and Problem Formulation}
\label{sec:system_model}

\begin{figure}[t]
    \centering
    \includegraphics[width=0.9\linewidth]{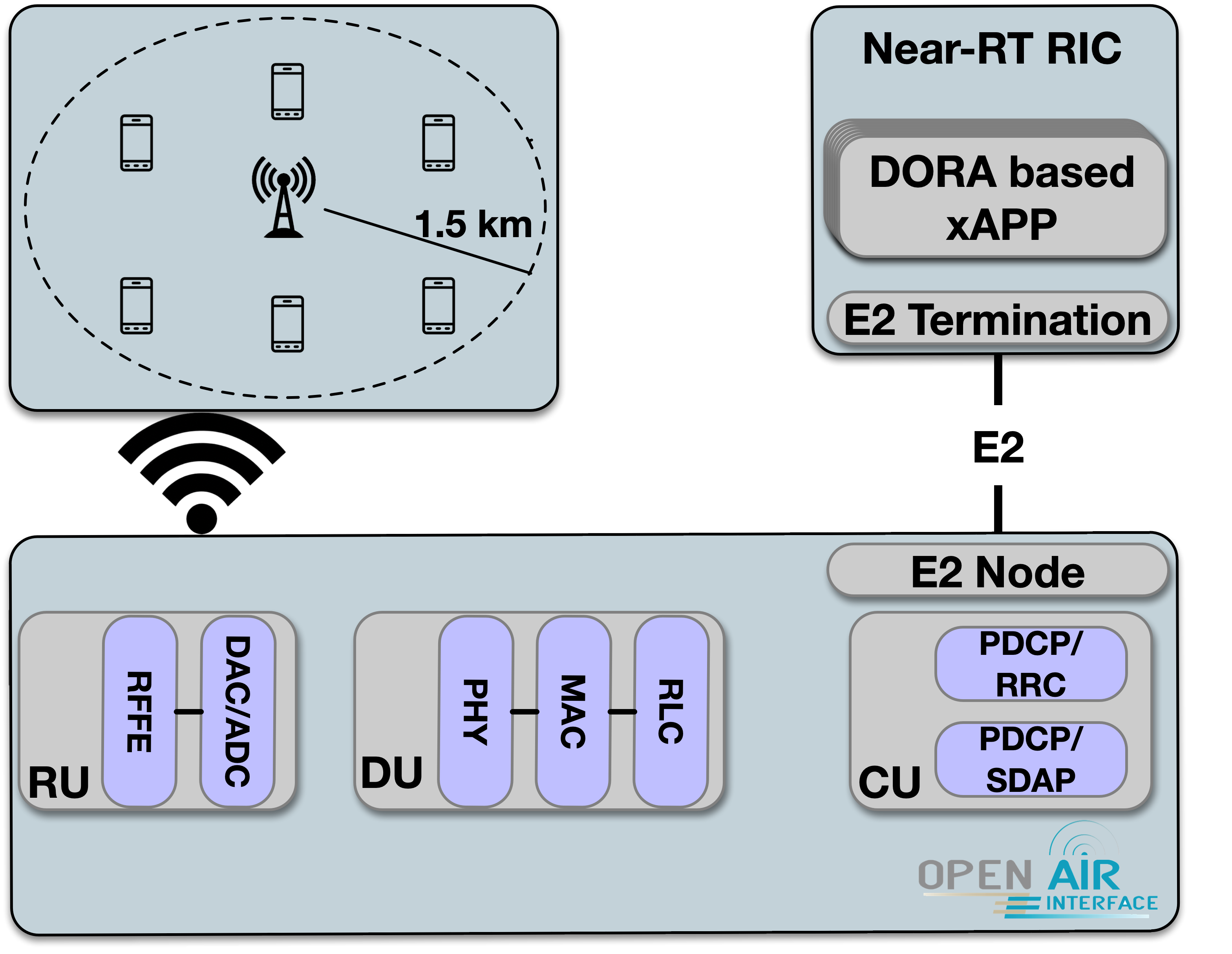} %
    \vspace{2pt} 
    \caption{DORA framework and E2-like integration with OAI. The near-RT RIC hosts the DORA xApp; control flows over an E2 interface to the E2 node (CU/DU/RU), enabling slice-level PRB decisions while preserving a 7.2 functional split.}
    \label{fig:dora_architecture}
\end{figure}

We consider a single 5G base station (gNB) serving three standardized network slices: 
Ultra-Reliable Low-Latency Communications (URLLC), enhanced Mobile Broadband (eMBB), 
and massive Machine-Type Communications (mMTC). The total downlink bandwidth is divided 
into $N_{\mathrm{PRB}} = 106$ Physical Resource Blocks (PRBs) per transmission time interval (TTI). 
The gNB dynamically allocates PRBs among slices to satisfy heterogeneous service-level 
agreements (SLAs) under variable channel and traffic conditions.

\subsection{Network Model}
The system consists of $N_U=2$ URLLC user equipments (UEs), $N_E=2$ eMBB UEs, and $N_M=10$ mMTC devices, with URLLC and eMBB UEs uniformly distributed in a $1500\,\text{m}\times1500\,\text{m}$ area and the gNB located at the center. URLLC UEs follow a random-walk mobility model with Bernoulli packet arrivals ($p=0.9$), packet sizes $\mathcal{U}[1500,4000]$~kbits, and a latency SLA of $400$\,ms. eMBB UEs also follow a random-walk mobility model, with stochastic throughput requests drawn from a normal distribution having an episode-mean of $7$\,Mbps and a standard deviation of $0.7$\,Mbps. mMTC devices are modeled as static terminals whose traffic is reservation-based; each device consumes $5$ PRBs from the shared pool, but since their channels are not explicitly simulated, pathloss and distance effects are not considered. All results focus on downlink performance under a unified PRB budget.

\subsection{Channel and Pathloss Model}
We adopt an OpenAirInterface-inspired offset free-space path loss (FSPL) model, 
which matches physical-layer measurements in our experimental setup. 
For a device located at distance $d$ from the gNB, the pathloss in dB is:
\begin{equation}
    \mathrm{PL}_{\mathrm{dB}} = 20\log_{10}(d) + 20\log_{10}(f) + 32.44 + C_{\mathrm{off}},
\end{equation}
where $f = 3.5\,\mathrm{GHz}$ is the carrier frequency and $C_{\mathrm{off}} = 83.84\,\mathrm{dB}$ 
is the calibration offset. Pathloss values are normalized into $[0,1]$ for inclusion in 
the RL observation vector. Throughput for a given $(\mathrm{PL}, N_{\mathrm{PRB}})$ pair 
is retrieved from an OAI-derived lookup table, which was generated 
from past throughput experiments conducted in OAI.

\subsection{Observation and Action Spaces}
At each step, the system state is represented by a continuous observation vector of size $3N$, 
where $N = N_U + N_E$ denotes the total number of URLLC and eMBB UEs:
\[
    \mathbf{o} = \{ s_i, r_i, p_i \}_{i=1}^{N},
\]

where $s_i \in \{0,1\}$ denotes the slice type (URLLC or eMBB), 
$r_i \in [0,1]$ is the normalized traffic demand 
(packet size for URLLC, requested throughput for eMBB), 
and $p_i \in [0,1]$ is the normalized pathloss. 
mMTC devices are excluded from per-device observations and handled slice-wise.

The action space is discrete and consists of $3003$ valid actions, 
each mapping to a PRB allocation tuple:
\[
    a = \left( \mathrm{PRB}_{\mathrm{URLLC}}, \mathrm{PRB}_{\mathrm{eMBB}}, \mathrm{PRB}_{\mathrm{mMTC}} \right),
\]
subject to:
\begin{align}{0}
\mathrm{PRB}_{\mathrm{URLLC}}  + \mathrm{PRB}_{\mathrm{eMBB}} 
+ \mathrm{PRB}_{\mathrm{mMTC}} = 106.
\label{eq:prb_constraints}
\end{align}

\subsection{Reward Function}
We formulate the PRB allocation problem as a discrete control Markov Decision Process (MDP), 
where the objective is to maximize a weighted sum of slice-level rewards:
\begin{equation}
    R = w_U R_U + w_E R_E + w_M R_M,
\end{equation}
with weights $(w_U, w_E, w_M) = (0.5,\,0.4,\,0.1)$. 
The slice-specific rewards are defined as:
\begin{align}
    R_U &= \max\left(-1, \min\left(0, \frac{\bar{L} - L_{\mathrm{tar}}}{L_{\mathrm{tar}}}\right)\right), \\
    R_E &= \max\left(-1, \min\left(0, \frac{T_{\mathrm{tar}} - \bar{T}}{T_{\mathrm{tar}}}\right)\right), \\
    R_M &= -1 + \frac{N_{\mathrm{serv}}}{N_{\mathrm{tot}}},
\end{align}
where $\bar{L}$ is the average URLLC latency, 
$L_{\mathrm{tar}} = 400\,\mathrm{ms}$ is the URLLC SLA, 
$\bar{T}$ is the average eMBB throughput, 
$T_{\mathrm{tar}} = 7\,\mathrm{Mbps}$ is the target eMBB SLA corresponding to the user-requested mean throughput, 
$N_{\mathrm{serv}}$ is the number of serviced mMTC devices, 
and $N_{\mathrm{tot}}$ is the total number of mMTC devices.

\section{DORA Framework}
\label{sec:dora}

\subsection{Architectural Overview and O-RAN Integration}
DORA is designed for seamless integration into O-RAN-compliant architectures, as illustrated in 
Fig.~\ref{fig:dora_architecture}. In O-RAN, the Near-Real-Time RIC (Near-RT RIC) hosts xApps responsible 
for adaptive RAN control, interfacing with E2 nodes, including the Central Unit (CU), Distributed Unit (DU), 
and Radio Unit (RU), via the standardized E2 interface. While O-RAN defines a complete E2 Application Protocol 
(E2AP) stack for exchanging control and telemetry messages, DORA adopts a Python-based \emph{E2-like} control architecture 
for lightweight deployment. Instead of relying on the full E2AP stack, this control architecture integrates directly with the OAI RAN stack via 
memory-mapped files and internal APIs. This enables real-time telemetry exchange, such as pathloss 
measurements and traffic demand profiles, while preserving a clear logical separation between control and 
data planes. Our deployment assumes an O-RAN-compliant 7.2 functional split, where the Radio Unit (RU) handles the 
RF front-end (RFFE) and DAC/ADC conversion, providing analog-to-digital and digital-to-analog processing 
for the radio signals. The Distributed Unit (DU) implements the PHY, MAC, and RLC layers, where the PHY 
performs modulation, coding, and channel estimation, the MAC manages scheduling and PRB allocation, and 
the RLC ensures reliable segmentation and reassembly of data. The Central Unit (CU) hosts two functional 
entities: the PDCP/RRC block, which performs ciphering, integrity protection, and connection control, and 
the PDCP/SDAP block, which handles QoS flow mapping and packet routing.

Within this O-RAN-inspired architecture, DORA employs a reinforcement learning (RL) agent hosted in the Near- RT RIC 
to optimize slice-level resource allocation dynamically. The agent observes network states, selects PRB 
allocation actions from a discrete set, and receives feedback via per-slice KPIs. To ensure real-world fidelity 
and scalability, DORA adopts a hybrid offline-online learning pipeline:
\begin{itemize}
    \item \textbf{Offline Training}: Accelerates convergence by pretraining the RL agent using an 
    OAI-derived throughput lookup table.
    \item \textbf{Online Evaluation}: Evaluates the trained policy under stochastic PHY/MAC realizations 
    through closed-loop interaction with OAI in real time.
\end{itemize}

\subsection{Offline Pretraining with OAI-Derived Lookup Tables}
Training RL agents directly in OAI is computationally prohibitive due to the need for high-fidelity 
PHY/MAC modeling. To mitigate this, DORA exploits a throughput lookup table,  indexed by $(\mathrm{PL}_{\mathrm{dB}}, N_{\mathrm{PRB}})$ pairs. 
This table is generated through controlled OAI simulation runs across representative pathloss and PRB configurations.  
During pretraining, the agent queries this table to estimate slice-level throughput, replacing 
on-the-fly channel emulation with table lookups. This strategy offers two advantages:  
(1) the agent experiences large-scale variability in traffic and channel conditions without 
incurring real-time simulation costs, and  
(2) policies converge significantly faster before deployment.

\subsection{Intra-Slice Scheduling}
While DORA operates at the slice level, intra-slice scheduling follows a round-robin policy 
among active URLLC and eMBB UEs. Given a slice-level allocation 
$(\mathrm{PRB}_{\mathrm{URLLC}}, \mathrm{PRB}_{\mathrm{eMBB}})$, PRBs are assigned one at a time 
to active UEs in a cyclic manner. For mMTC, per-device scheduling is not performed; instead, 
the number of serviced devices is given by:
\[
    N_{\mathrm{serv}} = \left\lfloor \frac{\mathrm{PRB}_{\mathrm{mMTC}}}{5} \right\rfloor.
\]
This separation between slice-level RL control and deterministic intra-slice policies simplifies 
the state and action spaces, allowing DORA to scale efficiently.

\section{Baseline Algorithms}
\label{sec:baselines}

To evaluate the effectiveness of the proposed DORA framework, we compare it against 
three representative non-learning baseline policies widely used in multi-slice PRB allocation. 
All baselines operate at the slice level and share the same intra-slice round-robin scheduler 
described in Section~\ref{sec:dora}. Performance differences therefore arise exclusively 
from the slice-level PRB allocation strategy.

\subsection{Hard Slicing (Fixed Ratios)}
The simplest baseline allocates a fixed proportion of the total PRBs to each slice, 
independent of instantaneous traffic demands or channel conditions. For example, a 
$40\%/40\%/20\%$ split is used for URLLC, eMBB, and mMTC, respectively:
\begin{align}
\mathrm{PRB}_{\mathrm{URLLC}} &= \left\lfloor 0.4\,N_{\mathrm{PRB}} \right\rfloor, \nonumber \\
\mathrm{PRB}_{\mathrm{eMBB}} &= \left\lfloor 0.4\,N_{\mathrm{PRB}} \right\rfloor, \nonumber \\
\mathrm{PRB}_{\mathrm{mMTC}} &= N_{\mathrm{PRB}} - 
    \mathrm{PRB}_{\mathrm{URLLC}} - \mathrm{PRB}_{\mathrm{eMBB}}.
\label{eq:prb_allocation}
\end{align}

This approach guarantees predictable per-slice bandwidth but lacks flexibility: unused PRBs 
cannot be reallocated when some slices are lightly loaded, resulting in underutilization or starvation under different demands.

\subsection{Priority-Based Allocation}
This baseline follows a sequential allocation strategy, prioritizing slices based on 
service-level agreements (SLAs). PRBs are allocated in the order 
URLLC $\rightarrow$ eMBB $\rightarrow$ mMTC:
\begin{enumerate}
    \item For each active URLLC UE, the PRBs required to meet the $400$\,ms latency SLA 
    are estimated via the OAI-derived throughput model; the total is allocated to URLLC, 
    capped by $N_{\mathrm{PRB}}$.
    \item Remaining PRBs are allocated to eMBB UEs to meet their requested throughout targets.
    \item Any residual PRBs are assigned to mMTC.
\end{enumerate}
While this strategy is more demand-aware than hard slicing, its offline SLA-based throughput 
estimation can mismatch online OAI dynamics, leading to over-allocation to URLLC and eMBB 
and frequent mMTC starvation.

\subsection{Fair Active-User Allocation}
This baseline emphasizes equity across active users rather than per-slice SLA targets.  
Up to $\lfloor N_{\mathrm{PRB}}/3 \rfloor$ PRBs are reserved for mMTC, capped by the number 
of devices serviced ($5$ PRBs per device). The remaining PRBs are distributed equally across 
all active URLLC and eMBB UEs:
\[
    \mathrm{PRB}_i = 
    \left\lfloor \frac{
    N_{\mathrm{PRB}} - \mathrm{PRB}_{\mathrm{mMTC}} }{
    N_U + N_E } \right\rfloor,\quad
    i \in \{\text{URLLC, eMBB}\}.
\]
Within each slice, PRBs are assigned round-robin to active UEs. While this policy improves 
inter-slice fairness, it ignores instantaneous traffic demands and can exacerbate URLLC 
latency violations and eMBB throughput deficits under high-load regimes.

\section{Experimental Setup}
\label{sec:experimental_setup}

To evaluate the performance of the proposed DORA framework, we integrate the 
offline-online pipeline with the OpenAirInterface (OAI) simulator to ensure 
realistic PHY/MAC dynamics. All experiments are designed to capture trade-offs 
between latency, throughput, and device coverage under resource scarcity.

\subsection{Simulation Environment}
We consider a single gNB deployed at the center of a $1500\,\mathrm{m} \times 1500\,\mathrm{m}$ area, 
serving $N_U=2$ URLLC UEs, $N_E=2$ eMBB UEs, and $N_M=10$ mMTC devices. 
The total number of available downlink PRBs per transmission time interval (TTI) 
is fixed at $N_{\mathrm{PRB}}=106$. 
Key simulation parameters are summarized in Table~\ref{tab:sim_params}.

\begin{table}[h]
\centering
\caption{Simulation Parameters}
\label{tab:sim_params}
\begin{tabular}{l l}
\toprule
\textbf{Parameter}                & \textbf{Value} \\ \midrule
Simulation area                  & $1500 \,\mathrm{m} \times 1500\,\mathrm{m}$ \\
Carrier frequency               & $3.5$ GHz \\
Total PRBs per TTI              & $106$ \\
Episode length                  & $256$ steps \\
URLLC UEs                       & $2$ \\
eMBB UEs                        & $2$ \\
mMTC devices                    & $10$ \\
mMTC PRB demand                 & $5$ PRBs/device \\
Pathloss model                  & OAI-based FSPL + offset \\
Throughput model                & OAI lookup table + Shannon fallback \\
Efficiency factor ($\eta$)      & $\approx 0.7$ \\
\bottomrule
\end{tabular}
\end{table}

\subsection{Traffic Regimes}
To expose meaningful trade-offs among slices, we deliberately construct traffic 
regimes where aggregate demand routinely exceeds the $106$ available PRBs.  
This stress-test setting reflects realistic scenarios where no allocation policy 
can simultaneously satisfy all slice-level SLAs, forcing the agent to learn 
adaptive prioritization strategies.

\begin{itemize}
    \item \textbf{URLLC:} Bernoulli arrivals ($p=0.9$) with packet sizes 
    $\mathcal{U}[1500,4000]$~kbits; SLA latency target $L_{\mathrm{tar}}=400$\,ms.
    \item \textbf{eMBB:} Stochastic per-step throughput requests drawn from a normal distribution with a per-step mean of $7$\,Mbps and a standard deviation of $0.7$\,Mbps.
    \item \textbf{mMTC:} Reservation-based access; a device is serviced if assigned 
    at least $5$ PRBs.
\end{itemize}

\subsection{Offline Training and Online Evaluation}
Offline training leverages an OAI-derived throughput lookup table indexed by 
$(\mathrm{PL}_{\mathrm{dB}}, N_{\mathrm{PRB}})$ pairs, capturing measured PHY-layer 
behavior across representative scenarios.  
The pre-trained policy is subsequently evaluated online in OAI, which introduces:
\begin{enumerate}
    \item Stochastic fading and interference patterns,
    \item Measurement noise in SINR and pathloss estimates,
    \item Cross-slice contention effects absent from offline approximations.
\end{enumerate}

\subsection{Evaluation Metrics}
We evaluate DORA and baseline policies using slice-specific KPIs and full 
distributional metrics:
\begin{itemize}
    \item \textbf{URLLC:} SLA violation rate, cumulative distribution function (CDF) 
    of end-to-end latency.
    \item \textbf{eMBB:} Throughput CDF, SLA violation rate relative to 
    the requested target throughput.
    \item \textbf{mMTC:} Number of concurrently serviced devices per episode.
\end{itemize}
In addition, we analyze multi-slice trade-offs by jointly visualizing URLLC latency, 
eMBB throughput, and mMTC coverage distributions to highlight the adaptive behavior 
of DORA under congestion.

\section{Results and Discussion}
\label{sec:results}

In this section, we present a comprehensive performance evaluation of the DORA framework, which employs a PPO-based agent. We benchmark its performance against three non-learning baselines—Hard Slicing, Priority-based, and Fair Active-User allocation—as well as a \texttt{DQN}-based DRL agent. The evaluation is conducted in a resource-constrained environment where the aggregate traffic demand frequently exceeds the available PRBs, compelling each allocation strategy to make critical trade-offs among the conflicting Quality of Service (QoS) requirements of the URLLC, eMBB, and mMTC slices.

\subsection{URLLC Latency Performance}

\begin{figure}[t]
    \centering
    \includegraphics[width=0.95\linewidth]{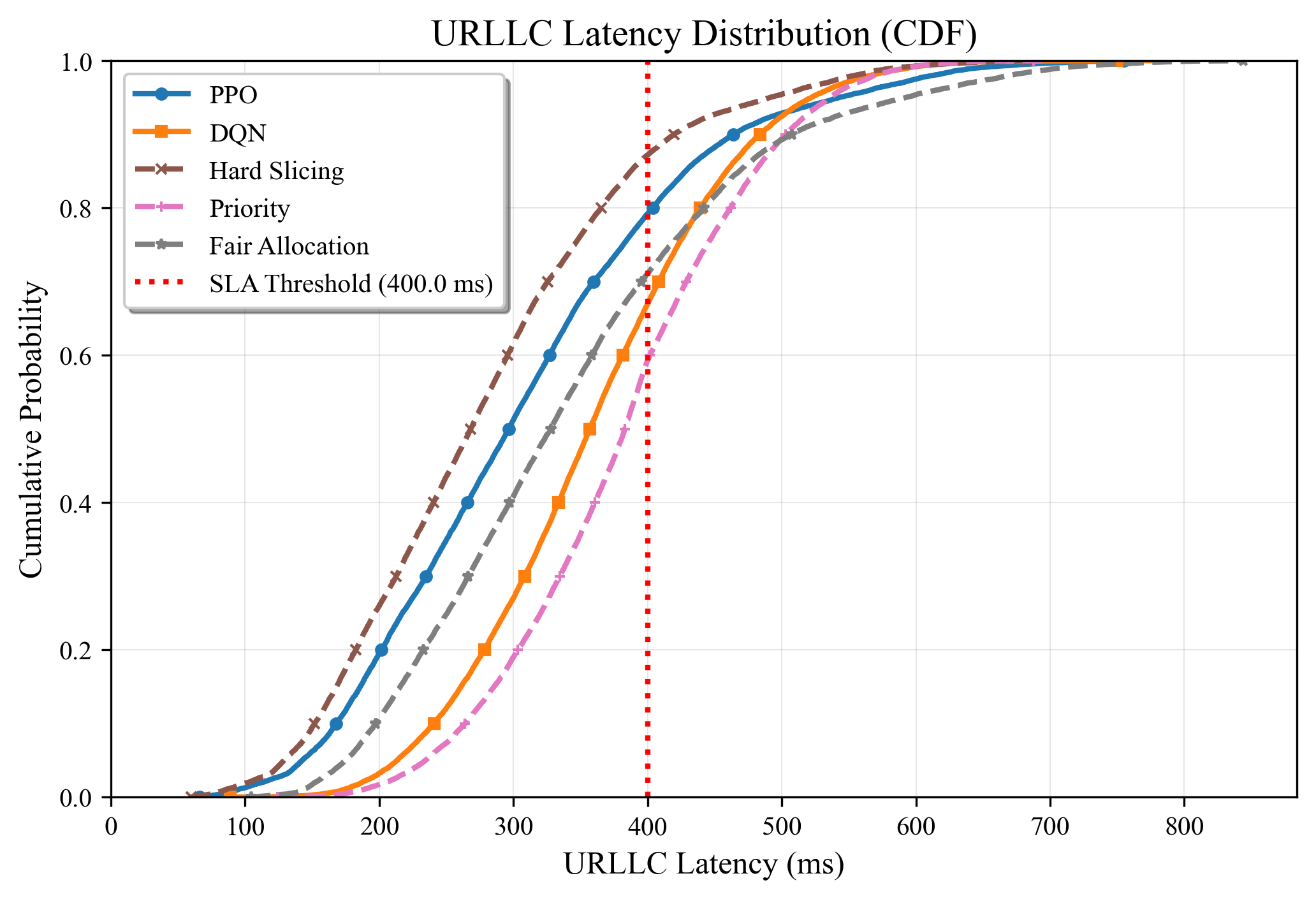}
    \caption{CDF of URLLC end-to-end latency. DORA (PPO) demonstrates strong SLA compliance, with its adaptive policy making slight compromises for improved overall system utility.}
    \label{fig:urllc_latency_cdf}
\end{figure}

The stringent latency requirement of the URLLC slice is a critical benchmark for any resource allocation strategy. Figure~\ref{fig:urllc_latency_cdf} illustrates the cumulative distribution function (CDF) of the end-to-end latency for URLLC users. At the strict 400\,ms SLA threshold, the Hard Slicing baseline exhibits the highest probability, indicating superior compliance due to its guaranteed, albeit inflexible, resource partitioning. DORA's PPO agent follows closely, demonstrating robust performance in meeting this critical SLA.

A key observation lies in the behavior of the CDF curves beyond the 400\,ms mark. The DORA agent's curve exhibits a steeper slope compared to the baselines. This suggests an intelligent and deliberate policy: the agent occasionally permits minor SLA violations for URLLC traffic to reallocate resources dynamically, thereby enhancing the overall system performance across all three slices. This calculated compromise is minimal; the PPO agent's CDF rapidly converges with the Hard Slicing benchmark, reaching the same cumulative probability value at approximately 440\,ms. This behavior highlights the agent's ability to not only respect the high priority of URLLC but also to minimize the magnitude of any deviation. In contrast, while the \texttt{DQN} agent also demonstrates adaptability, its overall latency performance is inferior to PPO, underscoring PPO's more effective decision-making in dynamic, high-stakes conditions.

\subsection{eMBB Throughput Analysis}

\begin{figure}[t]
    \centering
    \includegraphics[width=0.95\linewidth]{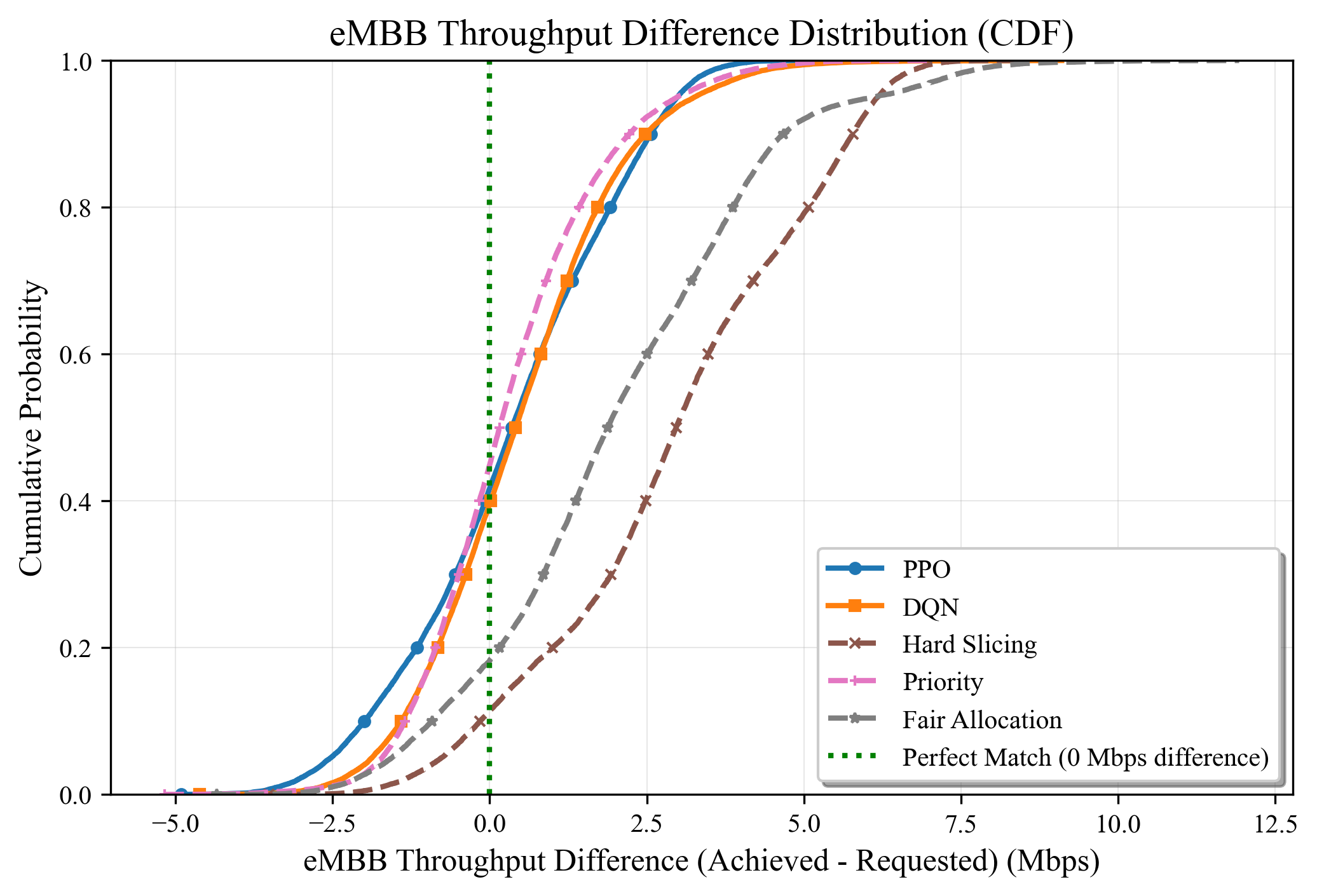}
    \caption{CDF of the difference between achieved and requested eMBB throughput. Negative values indicate SLA violations. DRL agents show strong adaptive performance comparable to the best-performing baselines.}
    \label{fig:embb_throughput_cdf}
\end{figure}

For the eMBB slice, performance is measured by the ability to meet stochastic user throughput demands, with an average request of 7\,Mbps. Figure~\ref{fig:embb_throughput_cdf} presents the CDF of the difference between the achieved and the requested throughput. The region to the left of the vertical green line (x=0) is of primary importance, as it represents SLA violations where users receive less throughput than demanded.

In this scenario, the Hard Slicing and Fair Allocation baselines perform strongly. Their static and user-agnostic allocation methods can be effective for maximizing throughput when demand is consistently high. The DRL agents, DORA (PPO) and \texttt{DQN}, deliver commendable and very similar performance, with \texttt{DQN} holding a slight edge. The steep slope of both DRL agents' CDFs signifies a high degree of adaptability and efficient resource management, as they actively work to minimize the deficit between requested and delivered throughput. This dynamic behavior contrasts with the more rigid nature of the non-learning baselines.

\subsection{mMTC Coverage Evaluation}

\begin{figure}[t]
    \centering
    \includegraphics[width=0.85\linewidth]{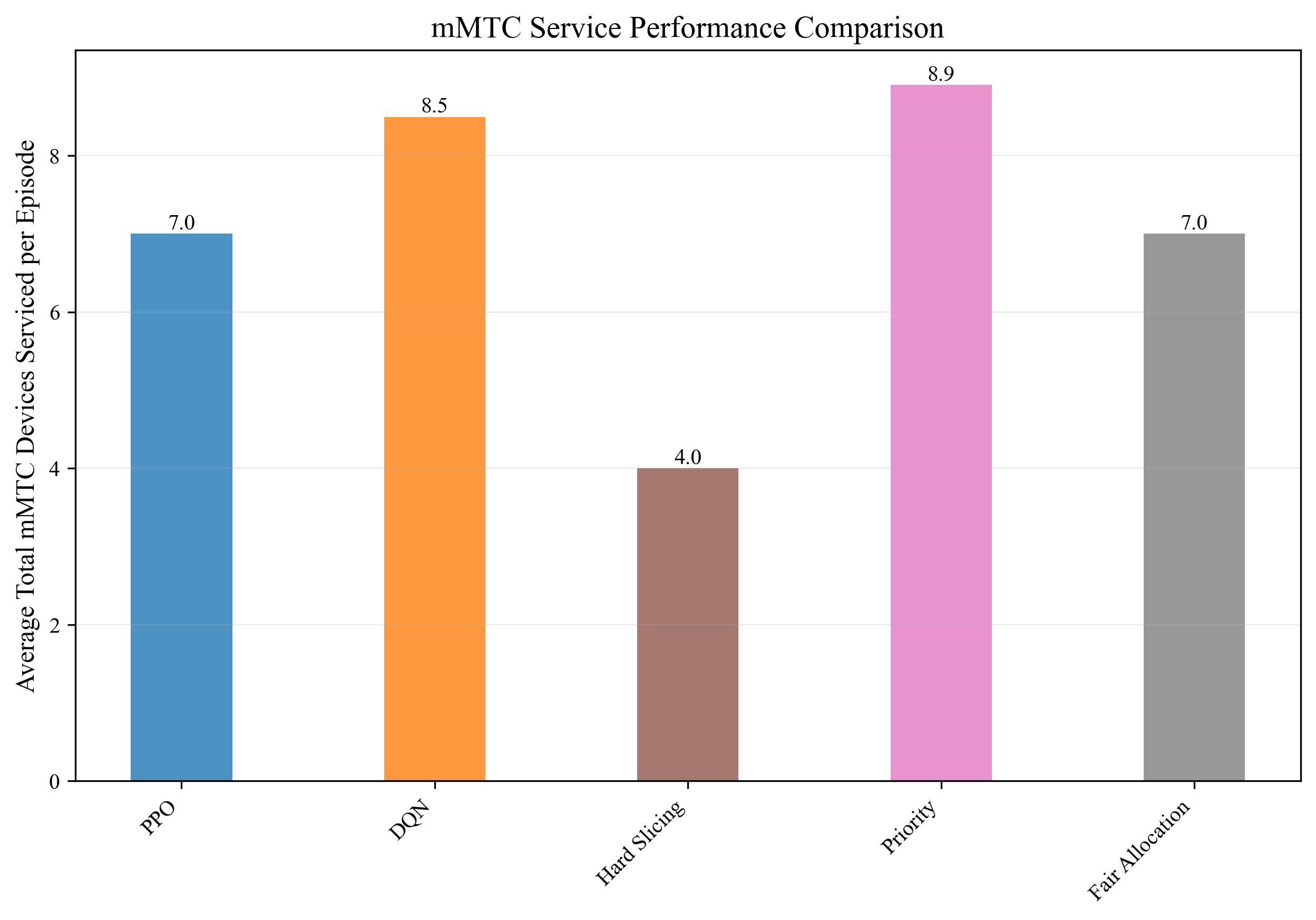}
    \caption{Average number of concurrently serviced mMTC devices. DORA achieves a balanced coverage, avoiding the starvation seen in the Hard Slicing baseline.}
    \label{fig:mmtc_serviced_barchart}
\end{figure}

The primary goal for the mMTC slice is to maximize the number of concurrently serviced devices. As shown in Figure~\ref{fig:mmtc_serviced_barchart}, the Priority-based and \texttt{DQN} agents achieve the best performance in this metric. The Priority-based approach allocates residual resources to mMTC, which can be substantial if URLLC and eMBB demands are met, while the \texttt{DQN} agent appears to have learned a policy that favors mMTC devices. DORA's PPO agent and the Fair Allocation baseline deliver comparable and effective coverage, servicing a significant number of devices. The Hard Slicing policy performs exceptionally poorly, servicing very few mMTC devices. This is a direct consequence of its static resource partitioning, which dedicates a fixed, large portion of PRBs to the URLLC and eMBB slices, irrespective of their instantaneous needs, thereby starving the lower-priority mMTC slice.

\subsection{Overall Performance and Synthesis}

The experimental results reveal that no single policy universally excels across all three slices, which underscores the inherent challenge of managing conflicting QoS requirements in 5G networks. However, the DRL-based frameworks, particularly DORA, demonstrate a superior ability to navigate these complex trade-offs. Unlike the rigid non-learning baselines, intelligent agents make dynamic, state-aware decisions to optimize for overall system utility. When comparing the DRL agents, DORA's PPO-based approach emerges as the more sophisticated and well-rounded solution. While the \texttt{DQN} agent shows strong performance for mMTC and eMBB, it does so at the expense of the high-priority URLLC slice. In contrast, DORA intelligently balances the objectives defined in the reward function, where mMTC carries the lowest weight ($w_M=0.1$). Its policy correctly prioritizes the critical URLLC latency SLA while still delivering competitive eMBB throughput and robust mMTC coverage. This demonstrates that DORA has successfully learned the nuanced, weighted objectives of the multi-slice environment, establishing it as a more reliable and effective framework for dynamic O-RAN resource allocation.


\bibliographystyle{IEEEtran}

\end{document}